\documentclass[twocolumn,aps,prl,superscriptaddress,showpacs,secnumroman,showkeys]{revtex4}
\usepackage{amsmath,bm}%
\usepackage{graphicx}%
\usepackage{epsf,epsfig,latexsym}
\begin{document}
\title{Experimental Determination of In-Medium Cluster Binding Energies and Mott Points in Nuclear Matter}  
\author{K. Hagel}
\affiliation{Cyclotron Institute, Texas A$\&$M University, College Station, Texas 77843}
\author{R. Wada}
\affiliation{Institute of Modern Physics HIRFL, Chinese Academy of Sciences, Lanzhou, 730000,China.}
\affiliation{Cyclotron Institute, Texas A$\&$M University, College Station, Texas 77843}
\author{L. Qin}
\affiliation{Cyclotron Institute, Texas A$\&$M University, College Station, Texas 77843}
\author{J. B. Natowitz}
\affiliation{Cyclotron Institute, Texas A$\&$M University, College Station, Texas 77843}
\author{S. Shlomo}
\affiliation{Cyclotron Institute, Texas A$\&$M University, College Station, Texas 77843}
\author{A. Bonasera}
\affiliation{Cyclotron Institute, Texas A$\&$M University, College Station, Texas 77843}
\affiliation{Laboratori Nazionali del Sud, INFN, via Santa Sofia, 62, 95123 Catania, Italy}
\author{G. R\"opke}
\affiliation{University of Rostock, FB Physik, Rostock, Germany }
\author{S. Typel}
\affiliation{GSI Helmholtzzentrum f\"ur Schwerionenforschung GmbH,Theorie, Planckstraße 1, D-64291 Darmstadt, Germany}
\author{Z. Chen}
\author{M. Huang}
\author{J. Wang}
\affiliation{Institute of Modern Physics HIRFL, Chinese Academy of Sciences, Lanzhou, 730000,China.}
\author{H. Zheng}
\affiliation{Cyclotron Institute, Texas A$\&$M University, College Station, Texas 77843}
\author{S. Kowalski}
\affiliation{Institute of Physics, Silesia University, Katowice, Poland.}
\author{C. Bottosso}
\author{M. Barbui}
\author{M. R. D. Rodrigues}
\author{K. Schmidt}
\affiliation{Cyclotron Institute, Texas A$\&$M University, College Station, Texas 77843}
\author{D. Fabris}
\affiliation{Dipartamento di Fisica dell~Universita di Padova and INFN Sezione di Padova, Padova, Italy}
\author{M. Lunardon}
\affiliation{Dipartamento di Fisica dell~Universita di Padova and INFN Sezione di Padova, Padova, Italy}
\author{S. Moretto}
\affiliation{Dipartamento di Fisica dell~Universita di Padova and INFN Sezione di Padova, Padova, Italy}
\author{G. Nebbia}
\affiliation{Dipartamento di Fisica dell~Universita di Padova and INFN Sezione di Padova, Padova, Italy}
\author{S. Pesente}
\affiliation{Dipartamento di Fisica dell~Universita di Padova and INFN Sezione di Padova, Padova, Italy}
\author{V. Rizzi}
\affiliation{Dipartamento di Fisica dell~Universita di Padova and INFN Sezione di Padova, Padova, Italy}
\author{G. Viesti}
\affiliation{Dipartamento di Fisica dell~Universita di Padova and INFN Sezione di Padova, Padova, Italy}
\author{M. Cinausero}
\affiliation{INFN Laboratori Nazionali di Legnaro, Legnaro, Italy}
\author{G. Prete}
\affiliation{INFN Laboratori Nazionali di Legnaro, Legnaro, Italy}
\author{T. Keutgen}
\author{Y. El Masri}
\affiliation{FNRS and IPN, Universit\'e Catholique de Louvain, B-1348 Louvain-Neuve, Belgium}
\author{Z. Majka}
\affiliation{Smoluchowski Institute of Physics, Jagiellonian University, Krakow, Poland}

\date{\today}

\begin{abstract}
In medium binding energies and Mott points for $d$, $t$, $^3$He and $\alpha$ 
clusters in low density nuclear matter have been determined at specific 
combinations of temperature and density in low density nuclear matter 
produced in  collisions of 47$A$ MeV $^{ 40}$Ar and $^{64}$Zn projectiles 
with $^{ 112}$Sn and $^{124}$Sn  target nuclei. The experimentally 
derived values of the in medium modified binding energies are in good 
agreement with recent theoretical predictions based upon the 
implementation of Pauli blocking effects in a quantum statistical approach. 
\end{abstract}



\maketitle
 
\section*{I. INTRODUCTION}

Simple nuclear statistical equilibrium models assume that properties of the 
species in equilibrium are the same as those of the isolated species. While 
this assumption is tenable at very low density, it is untenable at higher 
densities where in-medium effects lead to dissolution of the clusters and a 
transition to cluster-free nuclear matter. To deal with this intermediate 
density range, Typel \textit{et al.} \cite{typel10} have developed a 
quantum statistical approach which includes cluster correlations in the 
medium and interpolates between the exact low-density limit and the very 
successful relativistic mean field (RMF) approaches appropriate near the 
saturation density. The generalized RMF model developed attributes the 
decrease of the cluster 
fractions at high densities to a reduction of the cluster binding energies 
due to the Pauli blocking. This leads to the Mott effect of vanishing 
binding \cite{RMS}. Well-defined clusters appear only for densities below 
approximately 1/10 of the saturation density and get dissolved at higher 
densities. The maximum cluster density is reached around the Mott density. 
Because of the presence of strong correlations in the scattering state 
continuum that are effectively represented by one resonance, there is a 
non-vanishing cluster fraction above the Mott density \cite{schmidt90}. 
We report here the first experimental derivation of temperature and density 
dependent binding energies of $d$, $t$, $^3$He and $\alpha$ clusters, 
directly from experimental particle yields. Experimental values for Mott 
points for $d$, $t$, $^3$He and $\alpha$ clusters are in good agreement with 
the predictions made in reference \cite{typel10}.

\section{Experimental Techniques}
We reported in Refs. \cite{kowalski07} and \cite{natowitz10} that 
measurements of nucleon and light cluster emission from the participant 
matter which is produced in near Fermi energy heavy ion collisions could 
be employed to probe the EOS at low density and moderate temperatures 
where clustering is important. The NIMROD $4\pi$ multi-detector at 
Texas A\&M University has now been used to extend our measurements to 
higher densities. Cluster production in collisions of 47A MeV  $^{ 40}$Ar 
with $^{112,124}$Sn and $^{64}$Zn with $^{112,124}$Sn was studied. 
NIMROD consists of a 166 segment charged particle array set inside a 
neutron ball \cite{wuenschel09}.  The charged particle array is arranged 
in 12 rings of Si-CsI telescopes or single CsI detectors concentric 
around the beam axis. The CsI detectors are 1-10 cm thick Tl doped crystals 
read by photomultiplier tubes. A pulse shape discrimination method is 
employed to identify light particles in the CsI detectors. Neutron 
multiplicity is measured with the $4\pi$ neutron detector surrounding 
the charged particle array. The combined neutron and charged particle 
multiplicities were employed to select the most violent events for 
subsequent analysis. Further details on the detection system, energy 
calibrations and neutron ball efficiency may be found in 
reference \onlinecite{wuenschel09}. 

\section{Analysis}
The dynamics of the collision process allow us to probe the nature of the 
intermediate velocity ``nucleon-nucleon'' emission 
source \cite{mekjian78,hagel00,wada89,wada04}. Measurement of emission 
cross sections of nucleons and light clusters together with suitable 
application of a coalescence ansatz \cite{mekjian78} provides the means 
to probe the properties and evolution of the interaction region.  The 
techniques used have been detailed in several previous 
publications \cite{kowalski07,natowitz10,hagel00,wada89,wada04,qin08} and 
are described briefly below. A notable difference from 
references \onlinecite{kowalski07} and \onlinecite{natowitz10} is 
the method of density extraction.  This is discussed more extensively in 
the following. We emphasize that the event selection is on the more 
violent collisions. Cross section weighting favors mid-range impact parameters.

An initial estimation of emission multiplicities at each stage of the 
reaction was made by fitting the observed light particle spectra assuming 
contributions from three sources, a projectile-like fragment (PLF) source, 
an intermediate velocity (IV) source, and a target-like fragment (TLF) source. 
A reasonable reproduction of the observed spectra is achieved. 
Except for the most forward detector rings the data are dominated by 
particles associated with the IV and TLF sources. The IV source velocities 
are very close to 50\% of the beam velocity as seen in many other 
studies (\cite{mekjian78,hagel00,wada89,wada04} and references therein). 
The observed spectral slopes reflect the evolution dynamics of the 
source \cite{wada04,albergo85,kolomiets97}.   For further analysis, 
this IV source is most easily sampled at the intermediate angles where 
contributions from the other sources are minimized.  For the analysis 
of the evolution of the source we have selected the data in ring 9 
of the NIMROD detector. This ring covered an angular range in the 
laboratory of 38$^\circ$ to  52$^\circ$.  The results of the three-source 
fit analyses, as well as inspection of invariant velocity plots 
constructed for each ejectile and each system, indicate that this 
selection of angular range minimizes contributions from secondary 
evaporative decay of projectile like or target like sources \cite{qin08}. 

We treat the IV source as a nascent fireball created in the participant 
interaction zone.  The expansion and cooling of this zone leads to a 
correlated evolution of density and temperature which we probe using 
particle and cluster observables, yield, energy and angle.  As in the  
previous work \cite{kowalski07,natowitz10} we have employed double 
isotope yield ratios \cite{bauer95,zheng11} to characterize the 
temperature at a particular emission time. Model studies comparing 
Albergo model temperatures and densities to the known input values 
have shown the double isotope ratio temperatures to be relatively 
robust in this density range \cite{shlomo09}.  However the densities 
extracted using the Albergo model are useful only at the very lowest 
densities \cite{shlomo09}.  Both of these results are confirmed in 
the more extensive calculations of reference \onlinecite{typel10}. 
In this study we have employed a different means of density 
extraction, the thermal coalescence model of Mekjian \cite{hagel00,wada04}. 

To determine the coalescence parameter $P_{0}$, the radius in momentum space, 
from our data we have followed the Coulomb corrected coalescence model 
formalism of Awes \textit{et al.} \cite{awes81} and previously employed by us in 
reference \cite{mekjian78}.  In the laboratory frame the derived 
relationship between the observed cluster and proton differential cross 
sections is   

\begin{eqnarray}
\frac{d^2N(Z,N,E_A)}{dE_Ad\Omega} =& 
R_{np}^N\frac{A^{-1}}{N!Z!}\Big(\frac{\frac{4}{3}\pi P_0^3}{[2m^3(E-E_C)]
^{\frac{1}{2}}}\Big)^{A-1}&\nonumber\\
&\times\Big(\frac{d^2N(1,0,E)}{dEd\Omega}\Big)^A&
\label{eqnCoal}
\end{eqnarray}
where the double differential multiplicity for a cluster of mass number 
$A$ containing $Z$ protons and $N$ neutrons and having a Coulomb-corrected 
energy $E_A$, is related to the proton double differential multiplicity 
at the same Coulomb corrected energy per nucleon, $E-E_C$, where $E_{C}$
is the Coulomb barrier for proton emission.  
$R_{np}$ is the neutron to proton ratio.  Since within the framework of 
the coalescence model the yield ratios of two isotopes which differ by 
one neutron are determined by their binding  energies and the 
$n/p$ ratio in the coalescence volume, we 
have used the observed triton to $^{3}$He yield ratio to derive 
the $n/p$ ratio used in this analysis.    

In the Mekjian model thermal and chemical equilibrium determines coalescence 
yields of all species. Under these assumptions there is a direct relationship 
between the derived radius in momentum space and the volume of the emitting 
system.  In terms of the $P_0$ derived from Eq. (1) and assuming a spherical 
source
\begin{equation}
V=\Big(\Big(\frac{Z!N!A^3}{2^A}\Big)(2s+1)
e^{\frac{E_0}{T}}\Big)^{\frac{1}{(A-1)}}\frac{3h^3}{4\pi P_0^3}
\end{equation}
where $h$ is Plancks constant and $Z$, $N$, and $A$ are the same as in Eq. (1), 
$E_{0}$ is the binding energy and $s$ the spin of the emitted cluster and $T$ is 
the temperature.  Thus the volume can be derived from the observed $P_0$ and 
temperature values assuming a spherical shape in terms of the $P_0$ derived from 
Eq. (1).

Because our goal was to derive information on the density and temperature 
evolution of the emitting system, our analysis was not limited to 
determining an average $P_{0}$ value.   Instead, as in our previous 
studies \cite{kowalski07,natowitz10,hagel00}, results for $d$, $t$, $^3$He, 
and $^4$He, were derived as a function of $v_{\rm surf,}$ the velocity of the 
emerging particle at the nuclear surface, prior to Coulomb 
acceleration \cite{awes81}. From the relevant $P_{0}$ values we then 
determined volumes using Eq. (2). A comparison of these volumes indicated 
good agreement for $t$, $^{3}$He and $^{4}$He. The volumes  
derived from the deuteron data are typically somewhat smaller. This 
appears to reflect the fragility of the deuteron and its survival 
probability once formed \cite{cervesato92}.  For this reason we have 
used average volumes derived from the A=3 and 4 clusters to calculate 
the densities. Given that mass is removed from the system during the 
evolution, we determined the relevant masses for each volume by assuming 
that the initial mass of the source was that determined from the source 
fitting analysis and then determining the mass remaining at a 
given $v_{\rm surf}$ from the observed energy spectra. This is also an 
averaging process and ignores fluctuations. Densities were determined 
by dividing remaining masses by volumes.

\section{Results}
\subsection{Temperatures and Densities}
Inspection of the results for the four different systems studied revealed 
that the temperatures, densities for all systems are the same within 
statistical uncertainties. Therefore we have combined them to determine 
the values reported in this paper. 

We present, in Figure 1 the experimentally derived density and temperature 
evolution of the IV source.   Estimated  errors on the temperatures are 10\% 
below $\rho$ = 0.01 fm$^{-3}$increasing to 15\% at $\rho$ = 0.03 fm$^{-3}$.  
Estimated errors on the densities are 20\%.

In a recently submitted paper we reported equilibrium constants 
for $\alpha$ cluster formation as a function of temperature and 
density \cite{qin11}.  These equilibrium constants were then compared 
with those predicted by several different astrophysical equation of state 
models. Specifically we defined the equilibrium constants, $K_{c}$, for 
cluster formation in terms of density as

\begin{equation}
K_c(A,Z) = \rho_{(A,Z)}/[(\rho_p)^Z(\rho_n)^N]
\end{equation}
where $\rho_{(A,Z)}$ is the density of clusters of a specific mass 
number $A$ and atomic number $Z$, $N$ is the neutron number in the 
cluster and $\rho_p$ and $\rho_n$ are, respectively, the densities of 
free protons and neutrons. In the present work, we employ the observed 
temperature and density dependence of these equilibrium constants to 
extract the in medium modifications of the cluster binding energies and 
determine Mott points.  

\begin{figure}
\epsfig{file=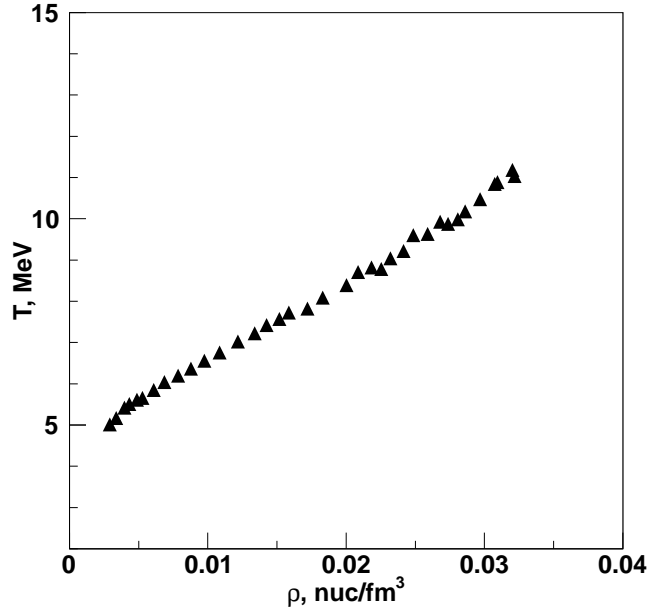,width=9.2cm,angle=0}
 \caption{Temperatures and densities sampled by the expanding IV source. }
\label{fig1}
\end{figure}

\section{Derivation of In Medium Binding Energies}
Our departure point for extraction of the medium modified cluster binding 
energies is the chemical equilibrium expression relating the density of a 
cluster of mass number $A$ and atomic number $Z$ to the densities of 
neutrons and protons in the same volume $V$ \cite{albergo85}.  

\begin{eqnarray}
\label{eq4}
\rho(A,Z) = \frac{N(A,Z)}{V} =
 &\frac{A^{\frac{3}{2}}\lambda_{T}^{3(A-1)}\omega(A,Z)}
{(2s_{p}+1)^Z(2s_{n}+1)^{A-Z}}&\nonumber\\
&\times\rho^Z_{p}\rho^{A-Z}_{n}\exp{\frac{B(A,Z)}{T}}
\end{eqnarray}

In this expression, $\lambda_T=\frac{h}{(2\pi m_0T)^{1/2}}$ is the thermal wavelength of a nucleon, 
s$_{p}$ and s$_{n}$ are the proton and neutron spins, $T$ is the temperature 
and $B(A,Z)$ is the cluster binding energy.  The term $\omega(A,Z)$ is 
the internal partition function of the cluster, taken here to 1 for 
the $Z=1$ and $Z =2$ clusters considered. 

Minich \textit{et al.} \cite{minich82,hirsch84} used a related yield 
expression to analyze intermediate mass fragment yields in 
multi-fragmentation experiments but added both a mixing entropy term and 
a surface entropy term,  as initially proposed by Fisher \cite{fisher66}. 
This latter term leads to a power law behavior of the mass distribution 
at the critical point and its ramifications have been widely 
explored \cite{elliott03,elliott00,huang10}. Neither of these entropy 
contributions is explicitly included in the Albergo 
formulation \cite{albergo85}. In the present analysis of the experimental 
data we include a mixing entropy term in the free energy. This term has 
the form 
\begin{equation}
\Delta F = T ( Z\ln(Z/A) + N\ln(N/A))
\end{equation}
where once again $Z$, $N$ and $A$ are those of the cluster being 
formed \cite{minich82}.  As mixing is a spontaneous process the free 
energy of mixing is negative and therefore favors the cluster formation. 
We do not include a Fisher term. Our reasoning for this is that, without 
additional corrections, the Fisher term as normally formulated and applied 
to larger clusters whose properties are very similar to the bulk properties, 
is not applicable to the yields of the very small clusters, $A \leq$ 4, 
which we are treating. We base this conclusion on the results of molecular 
dynamics studies of the cluster size dependence of the surface 
energy \cite{hale96,kiefer00,curtin83} and the binding energies per nucleon 
of the competing $Z=1,2$ species being significantly different from the bulk.

\begin{figure}
\epsfig{file=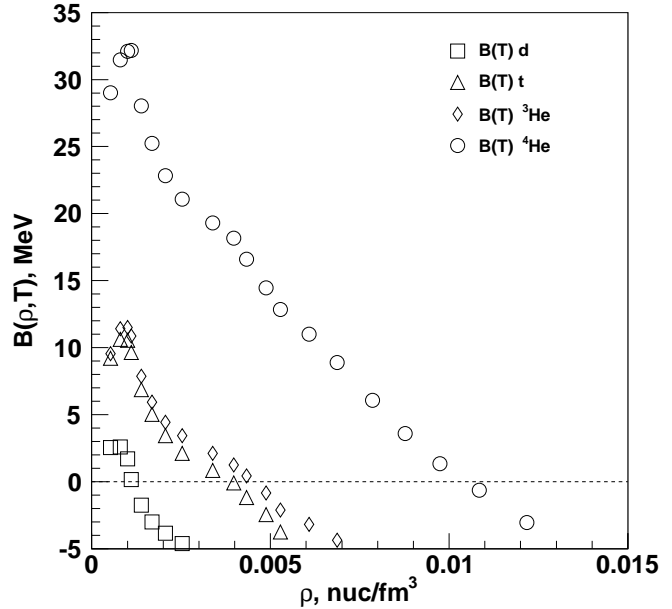,width=9.2cm,angle=0}
 \caption{In medium binding energies derived from the experiments as a 
function of density.  $T$ and $\rho$ are changing in a correlated fashion. 
(See text.)}
\label{fig2}
\end{figure}

Thus, rearranging Eq. (4) , substituting $K_{c}$ from Eq. (3) and 
taking the logarithm of each side we can write a general expression for 
each cluster, 
\begin{equation}
\ln[K_{c}/ C(T)]   =  B/T - Z\ln (Z/A) - N\ln (N/A)
\end{equation}
where $C(T)$ includes all terms on the right hand side of Eq. (\ref{eq4}) 
except the exponential term. Using the experimentally determined equilibrium 
constants and temperatures we then solve this expression to obtain the 
apparent binding energies, $B(\rho,T)$, of the clusters for the 
different temperatures and densities sampled in the experiments.  The 
binding energies extracted for $d$, $t$, $^3$He and $\alpha$ clusters 
decrease monotonically with increasing density as shown in Fig.~\ref{fig2}.   

\section{Mott Points}
By definition, a Mott point corresponds to a combination of density and 
temperature at which a cluster binding energy, $B(\rho,T)$,  is zero with 
respect to the surrounding medium. Since the observed temperatures and 
densities are correlated in our experiment (see Figure 1) each point in 
Figure 2 at which the experimentally derived binding energy is zero 
corresponds to a particular combination of density and temperature. 
Thus, with the present data, we are able to extract a single Mott point 
for each cluster. In Figure 3 we present the values of the Mott 
temperatures and densities and compare them with the loci of the values 
of  medium modified binding energies predicted by 
Typel \textit{et al.} \cite{typel10} using the thermodynamic Green function method. 
Such a Mott line was also calculated in reference \cite{stein95} while
the contribution of correlations was considered without discriminating among 
different clusters.  This approach makes explicit use of an effective 
nucleon-nucleon interaction to account for medium effects on the 
cluster properties \cite{roepke83}.  We see that the agreement between 
the predictions and the experimental results is quite good.

\begin{figure}
\epsfig{file=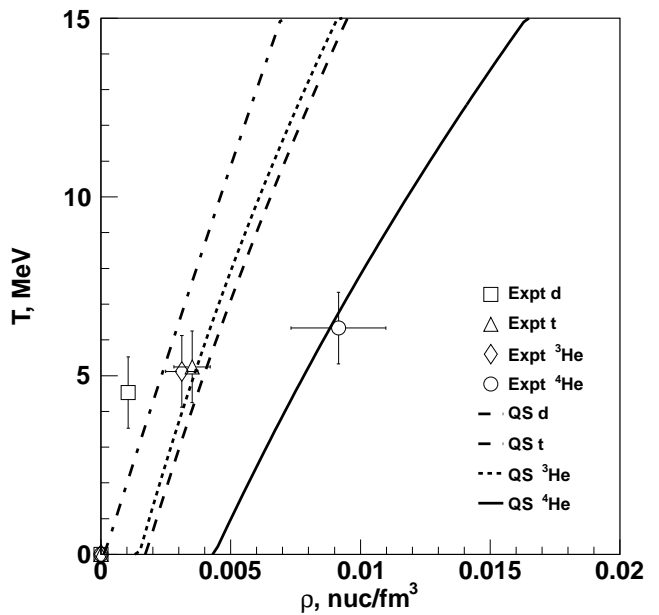,width=9.2cm,angle=0}
 \caption{Comparison of experimentally derived Mott point densities and 
temperatures with theoretical values. Symbols represent the experimental 
data.  Estimated errors on the temperatures are 10\% and on the 
densities 20\%.  Lines show polynomial fits to the Mott points presented 
in reference \cite{typel10}. }
\label{fig3}
\end{figure}
                               
\section{Summary and Conclusions}

We have presented a first experimental determination of in medium cluster 
binding energies and Mott points for $d$, $t$, $^3$He and $\alpha$ clusters 
produced in low density nuclear matter.  Our results are in good agreement 
with those predicted by a recent model which explicitly treats these 
quantities.  Inclusion of the in-medium  effects in astrophysical 
equations of state should improve the utility of those for modeling 
astrophysically interesting events. 

\section{Acknowledgements}
This work was supported by the United States Department of Energy under 
Grant \# DE-FG03- 93ER40773 and by The Robert A. Welch Foundation under 
Grant \# A0330.

\end{document}